\documentclass[twocolumn]{jpsj2} 
\usepackage{txfonts}

\title{Toward Identification of Order Parameters in Skutterudites\\
-- a Wonderland of Strong Correlation Physics --}

\author{
Yoshio
\textsc{Kuramoto}\thanks{E-mail address: kuramoto@cmpt.phys.tohoku.ac.jp} and 
Annam\'aria \textsc{Kiss}\thanks{E-mail address: amk@cmpt.phys.tohoku.ac.jp} 
}

\inst{Department of Physics, Tohoku University, Sendai, 980-8578}
\abst{
Current status is described toward identifying unconventional order parameters in filled skutterudites with unique ordering phenomena.  
The order parameters in PrFe$_4$P$_{12}$ and PrRu$_4$P$_{12}$ are discussed in relation to associated crystalline electric field (CEF) states and angular form factors. 
By phenomenological Landau analysis, it is shown that a scalar order model
explains most properties in both PrFe$_4$P$_{12}$ and PrRu$_4$P$_{12}$ with very different magnetic properties.  In particular, the highly anisotropic susceptibility induced by uniaxial pressure in PrFe$_4$P$_{12}$ is explained in terms of two types of 
couplings. 
In the case of SmRu$_4$P$_{12}$, 
the main order parameter at low field is identified as magnetic octupoles.  
A microscopic mechanism is proposed how the dipole and octupole degrees of freedom mix under the point group $T_h$ of skutterudites.
}

\kword{skutterudite, PrFe$_4$P$_{12}$, PrRu$_4$P$_{12}$, 
SmRu$_4$P$_{12}$, scalar order, hybridization, uniaxial pressure}

\begin{document}
\maketitle

\section{Introduction}

Filled skutterudite compounds provide an unprecedented framework where 
fundamental and long-standing problems in condensed-matter physics
all show up within the same crystal structure.  
In rare earth (R) skutterudites RT$_{4}$X$_{12}$ with T transition metals and X pnictogens, different combinations of constituent atoms lead to enormously rich variety of properties, such as itinerant-localized dichotomy and hidden electronic orders. 
At the present stage,  theoretical work needs to combine a phenomenological Landau-type approach and microscopic consideration for understanding the overall behavior.
In this paper, we focus on identification of unconventional order parameters in PrFe$_4$P$_{12}$,  PrRu$_4$P$_{12}$ and SmRu$_4$P$_{12}$.

PrFe$_{4}$P$_{12}$ undergoes a second-order phase transition at $T_{0}=6.5$K, which accompanies a typical structure in the specific heat, 
a sharp peak in the magnetic susceptibility, and a steep increase of the resistivity just below the transition \cite{aoki}.
A crystal-structure modulation with the wave vector ${\bf Q}=(1,0,0)$ was found below $T_{0}$ by X-ray diffraction experiments \cite{iwasa3}, which is attributed to the existence of staggered local electronic states of the Pr ions.
Early NMR \cite{kikuchi2} and elastic measurements \cite{nakanishi1} were interpreted in terms of antiferro ordering of $\Gamma_{3}$-type quadrupole moments.
However, the persistent isotropy of the magnetic susceptibility in the ordered phase cannot be explained by $\Gamma_{3}$ quadrupolar order. 
Furthermore, staggered dipoles are always parallel to the field direction both in neutron diffraction \cite{hao2, hao1} and NMR \cite{kikuchi1}.
Recently, careful analysis of the NMR results have shown that the local symmetry at the Pr sites is preserved in the ordered phase \cite{kikuchi1, sakai}. 
Furthermore, the continuous field-angle dependence of the transition temperature gives also evidence for the exclusion of $\Gamma_{3}$ quadrupolar order \cite{kiss06, sakakibara}.

In a recent paper \cite{kiss06}, we have proposed that the order parameter in 
PrFe$_4$P$_{12}$ is a staggered electronic order which does not break the local $T_h$ symmetry around each Pr site.  
We call this order a scalar order since it has
the $\Gamma_{1g}$ symmetry. 
It is found in ref.~\citen{kiss06} that the scalar order scenario can explain naturally the isotropic magnetic susceptibility in the ordered phase, the field angle dependence of the transition temperature and magnetization, and also the splitting pattern of the $^{31}$P NMR spectra.
In this paper,  we proceed to CEF theoretical description of the scalar order, and compare PrFe$_4$P$_{12}$ and PrRu$_4$P$_{12}$.
Furthermore, 
we explain not only the N\'{e}el-type anomalies of the magnetic susceptibility $\chi$ in PrFe$_4$P$_{12}$ near the phase transition, but also the huge anisotropy of $\chi$ induced under uniaxial pressure.  

As another prototype of mysterious orders, we take SmRu$_4$P$_{12}$, and 
analyze its CEF states.  
The conventional view point is that the CEF states are composed by linear combination of Hund's rule ground state with $J=5/2$.  In this case, 
the highest rank of multipoles in this manifold is $2J=5$, which is too small to distinguish between the $O_h$ and $T_h$ symmetries.  
Namely, the Stevens operator of sixth rank $O_6^t=O_6^2-O_6^6$, which makes the difference \cite{takegahara}, has zero matrix elements with $J=5/2$.
On the other hand, recent experimental results suggest mixing of dipole and octupole degrees of freedom \cite{mito,yoshizawa1}.
We analyze the wave functions in the CEF states taking higher order hybridization processes. 
It is found that the closeness of the $J=7/2$ excited state
above the $J=5/2$ ground state tends to compensate the small ratio of hybridization over excitation energy of $4f^6$ intermediate states.   

\section{CEF states and scalar orders in Pr skutterudites}

\subsection{Relevant CEF states}

In previous work\cite{otsuki}, we ascribed the main source of CEF splittings to covalent hybridization effects between $4f$ and ligand orbitals.  
The relevant point group $T_h$ causes mixing of two kinds of triplets $\Gamma_4$ and $\Gamma_5$ in the cubic case.  The hybridized triplets are called 
$\Gamma_4^{(1)}$ with larger weight from $\Gamma_4$, and 
$\Gamma_4^{(2)}$ with larger weight from $\Gamma_5$.
For later use, we shall give explicit form of these states in the case of $4f^2$ configuration with $J=L-S=5-1=4$.
\begin{align}
|\Gamma_1\rangle &=\sqrt{5/24}\left( |4\rangle +|-4\rangle  \right)
+\sqrt{7/12}|0\rangle, \label{Gamma1}
\\
|\Gamma_4; a \rangle &=\sqrt{1/2}\left( |4\rangle -|-4\rangle  \right), \\
|\Gamma_4; b \rangle &=\sqrt{1/8}|3\rangle+\sqrt{7/8}|-1\rangle, \\
|\Gamma_4; c \rangle &=\sqrt{1/8}|-3\rangle+\sqrt{7/8}|1\rangle, \\
|\Gamma_5; a \rangle &=\sqrt{1/2}\left( |2\rangle -|-2\rangle  \right), \\
|\Gamma_5; b \rangle &=\sqrt{7/8}|3\rangle-\sqrt{1/8}|-1\rangle, \\
|\Gamma_5; c \rangle &=\sqrt{7/8}|-3\rangle-\sqrt{1/8}|1\rangle, 
\end{align}
in terms of eigenstates of $J_z$.
Similarly, the double $\Gamma_3$ CEF states are given explicitly by
\begin{align}
|\Gamma_3; a \rangle &=\sqrt{7/24}\left( |4\rangle +|4\rangle  \right)
-\sqrt{5/12}|0\rangle, \\
|\Gamma_3; b \rangle &=\sqrt{1/2}\left( |2\rangle +|2\rangle  \right).
\label{Gamma4b}
\end{align}

The three representative Pr skutterudites, 
PrFe$_4$P$_{12}$, 
PrRu$_4$P$_{12}$, and
PrOs$_4$Sb$_{12}$ 
all have the singlet CEF ground state.  However, the first excited level is different from each other.  
Namely, PrFe$_4$P$_{12}$ has the low-lying $\Gamma_4^{(1)}$ triplet with strong van-Vleck susceptibility, and possibly the $\Gamma_{23}$ doublet, which goes over to $\Gamma_3$ doublet in the cubic symmetry.  
We suspect that these six levels are almost degenerate in the high-temperature phase of PrFe$_4$P$_{12}$.  As temperature becomes lower than $T_0$, one of the Pr sublattices has a singlet CEF ground state, while the other Pr appears to take the doublet.  This conjecture comes from neutron scattering of  PrFe$_4$P$_{12}$ where at least two inelastic transitions are visible in the ordered phase \cite{iwasa1}.

On the other hand, PrOs$_4$Sb$_{12}$ has the low-lying $\Gamma_4^{(2)}$ triplet which goes over to $\Gamma_5$ in the cubic point-group symmetry.  Hence the quadrupolar (van-Vleck) susceptibility is large in PrOs$_4$Sb$_{12}$.
Finally, the triplet in PrRu$_4$P$_{12}$ in the high-temperature phase appears to be a strong mixture of $\Gamma_4$ and $\Gamma_5$.  In the ordered phase, one of the Pr sublattices has the singlet CEF ground state, while the other has the crossing of singlet and $\Gamma_4^{(2)}$ triplet levels with decreasing temperature.  
Further decrease of temperature brings about the point group lower than $T_h$ as observed in the splitting of the triplet of the order of 1 K\cite{aoki2}.

\subsection{Angular form factors associated with the scalar order}

Multipolar interactions of rank four (hexadecapole) or rank six (hexacontatetrapole)
have a chance to bring an electronic order which keeps the original $T_h$ symmetry around each Pr site, but lead to  
A and B sublattices with different CEF ground states.
The shape of the Fermi surface with good nesting property should be responsible for the staggered AB sublattice structure \cite{harima,takimoto}.  
Since the scalar order accompanies a slight lattice distortion, it can be probed by X-ray diffraction \cite{iwasa3}.  
More detailed information should be obtained if
$4f$ form factors are probed by azimuthal scan in resonant X-ray scattering using the electric quadrupole ($E2$) channel \cite{kuramoto06}.

Since the fourth-rank tensor relevant to $E2$ scattering is very complicated,  we visualize the scalar order by deriving the simplest form factor that corresponds to a weighted average of the electron charge density.
Namely, we
utilize the integer $(J=4)$ value of the Pr$^{3+}$ configuration, and introduce a fictitious ``wave function":
\begin{align}
\psi_{\Gamma\alpha} (\Omega)= \langle \Omega|\Gamma,\alpha\rangle,
\end{align}
where $\Omega$ represents the solid angle specified by $(\theta,\phi)$ such that
$d\Omega =\sin\theta d\theta d\phi$.
Then $\psi_{\Gamma\alpha} (\Omega)$
can be derived in terms of spherical harmonics $Y_{4m}(\Omega)$ with use of eqs.(\ref{Gamma1}) to (\ref{Gamma4b}).
The angular form factor $\rho_{\Gamma J} (\Omega)$ associated with
a CEF level $\Gamma$ is defined by
\begin{align}
\rho_{\Gamma J} (\Omega) = \sum_\alpha w_\alpha
|\psi_{\Gamma\alpha} (\Omega)|^2,
\end{align}
where $w_\alpha$ is the weight factor of the component $\alpha$ in the CEF states.  
In the case of singlet $\Gamma_1$, we have $w_\alpha=1$, while in the case of doublet or triplet, we have $w_\alpha=1/2$ or $w_\alpha=1/3$, respectively.  
Since $|\Gamma,\alpha\rangle$ is not a single-particle state, 
and since spins are also involved,
the form factor 
$\rho_{\Gamma J} (\Omega)$ 
is not a charge density itself.
One may nevertheless gain good insight into spatial pattern of the scalar order by $\rho_{\Gamma J} (\Omega)$.

\begin{figure}
\centering
\includegraphics[width=0.22\textwidth]{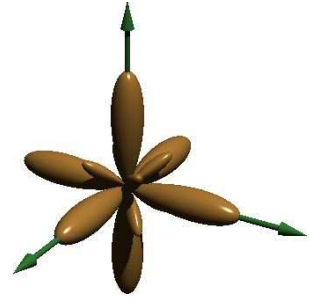}
\caption{
Angular form factor of $\Gamma_{1}$ singlet CEF state.
}
\label{Gamma_1}
\end{figure}

Figure \ref{Gamma_1} illustrates the angular form factor
associated with the CEF singlet state $\Gamma_1$.   
The distance of a point on the surface from the origin represents $\rho_{\Gamma J} (\Omega)$ for each solid angle.
Clearly the cubic symmetry is preserved in the form factor, which is understood as a superposition of a constant, a hexadecapole $\hat{x}^4+\hat{y}^4+\hat{z}^4-3/5$ with $\hat{\mib r}=(\hat{x},\hat{y},\hat{z})$ being a unit vector, and a hexacontatetrapole $(\hat{x}^2-\hat{y}^2)(\hat{y}^2-\hat{z}^2)(\hat{z}^2-\hat{x}^2)$.
In the case of the multiplet, on the other hand, each wave function breaks the cubic symmetry.  
However, summation over equally distributed degenerate states recovers
the pattern consistent with the cubic symmetry.  

\begin{figure}
\centering
\includegraphics[width=0.22\textwidth]{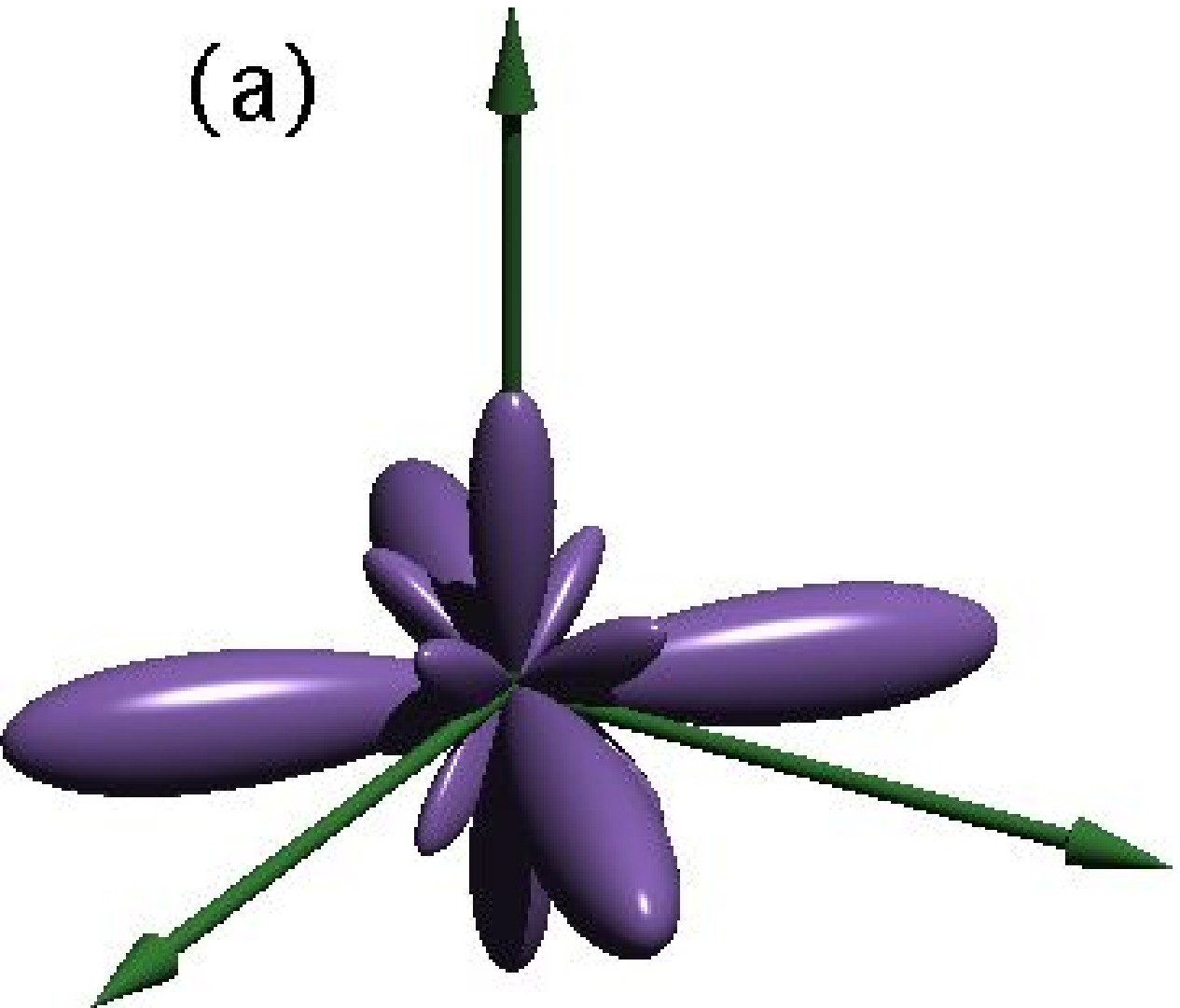}\hspace*{0.3cm}
\includegraphics[width=0.2\textwidth]{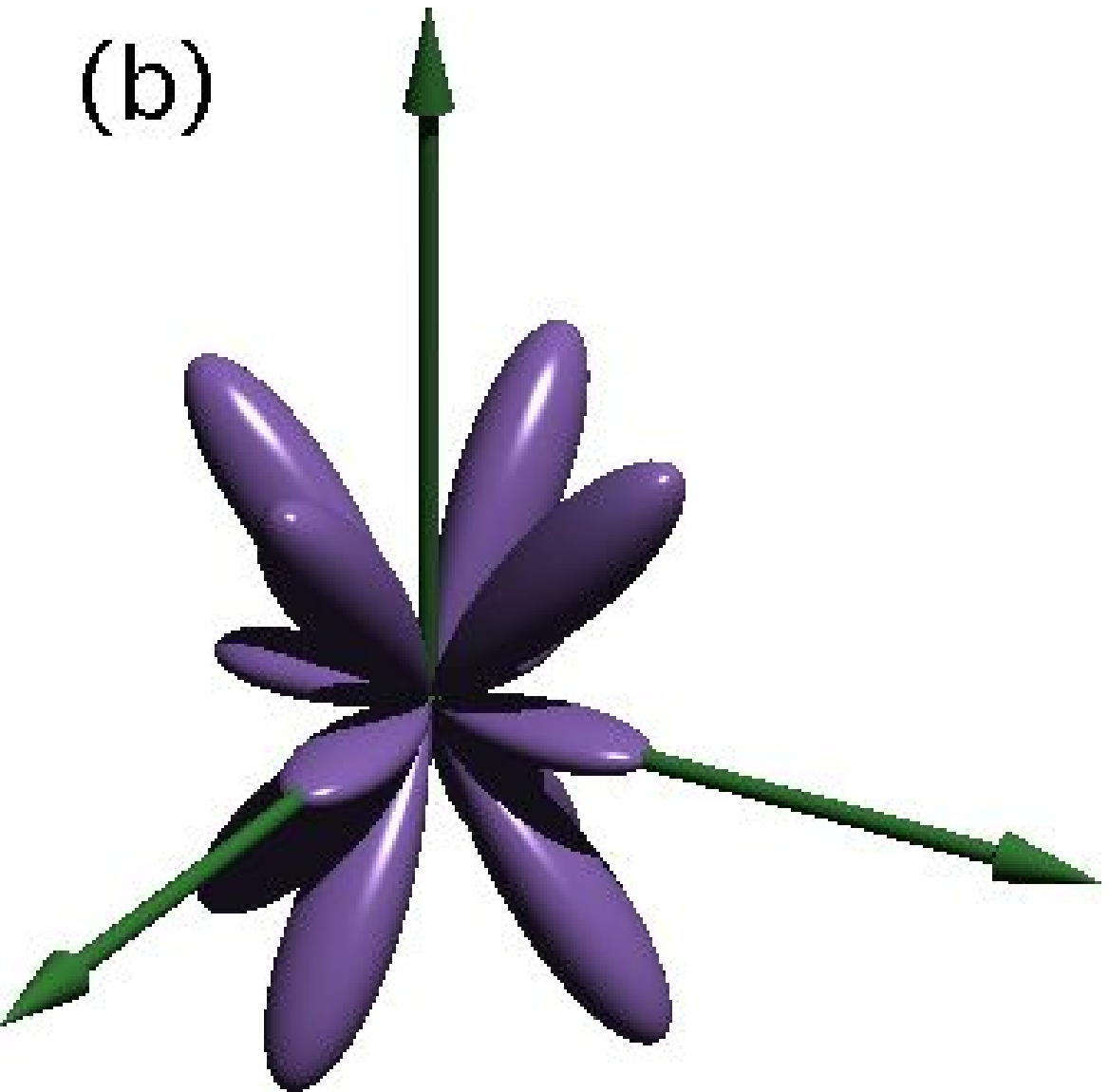}\\
\includegraphics[width=0.2\textwidth]{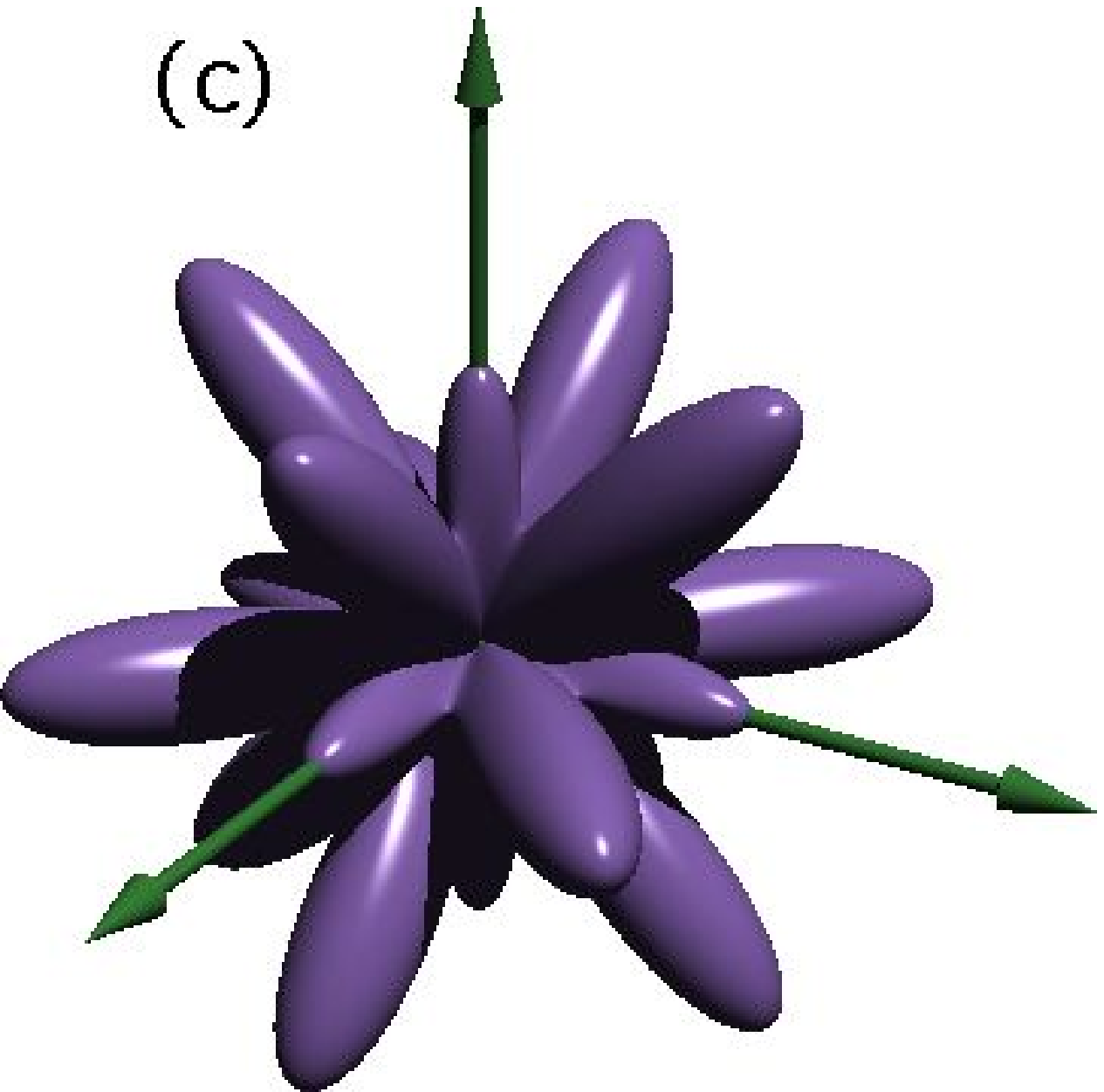}
\caption{
Angular form factor of $\Gamma_{3}$ doublet states: 
(a) $|\Gamma_{3};a\rangle$ state; (b) $|\Gamma_{3};b\rangle$  state; (c) average over the both states.
}\label{Gamma_3}
\end{figure}

\begin{figure}
\centering
\includegraphics[width=0.18\textwidth]{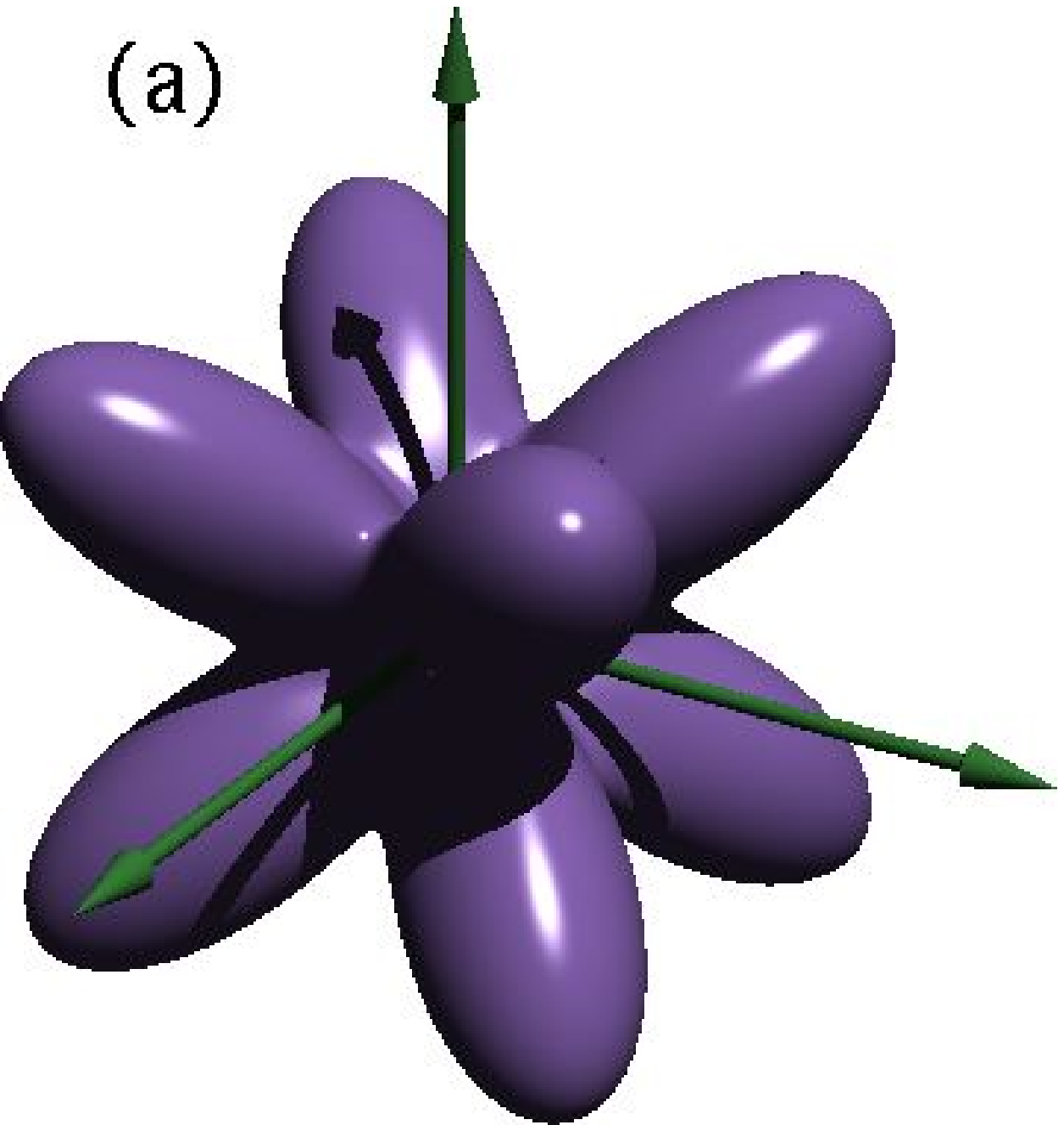}\hspace*{0.6cm}
\includegraphics[width=0.18\textwidth]{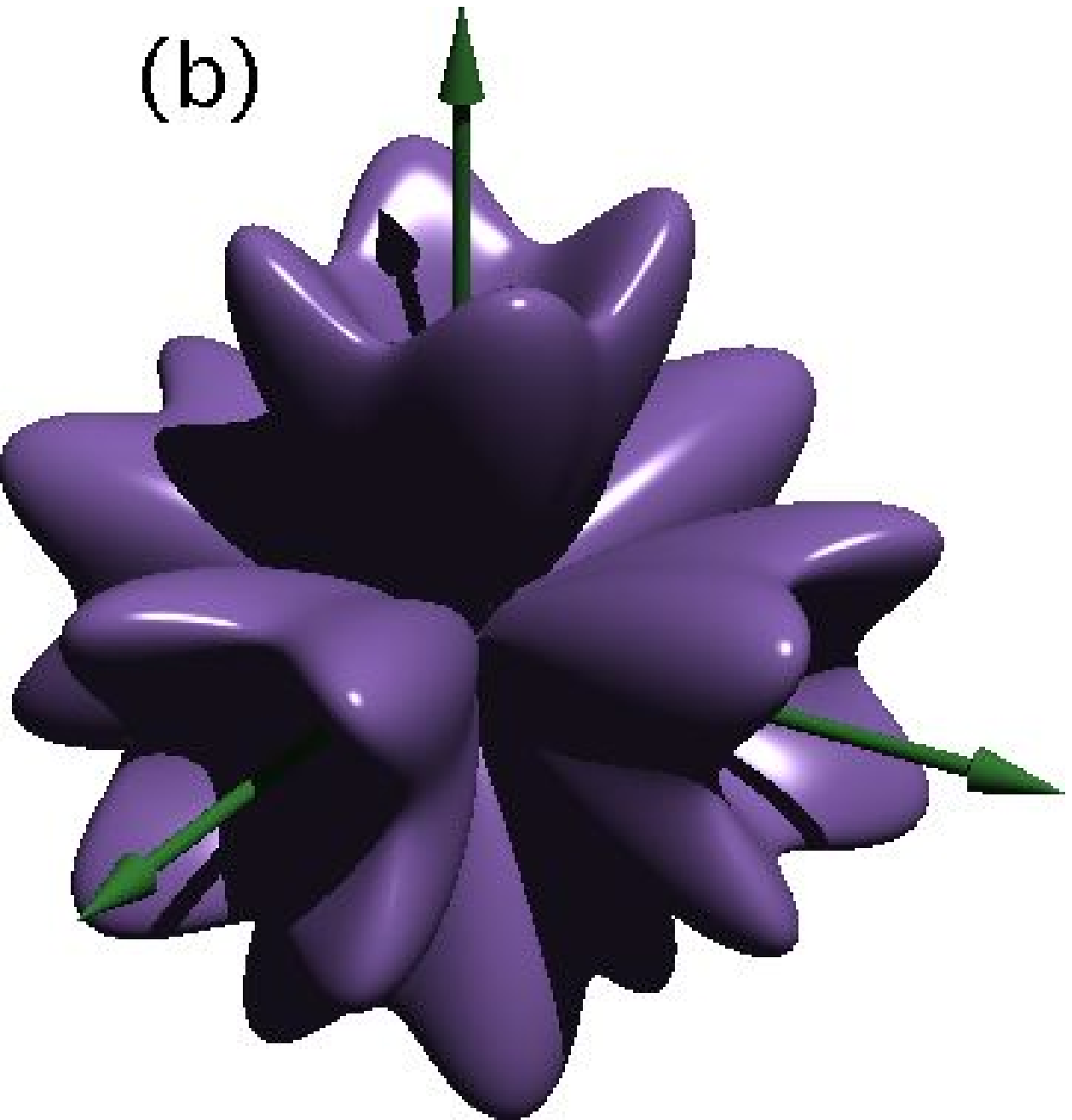}
\caption{
Angular form factor of triplet states averaged over three components:
(a) $\Gamma_{5}$ state;
 (b) $\Gamma_{4}$ state.
}\label{Gamma_45}
\end{figure}

Figure \ref{Gamma_3} shows the form factor of each $\Gamma_3$ state (a) and (b), and the average of the two in (c).  
Figure \ref{Gamma_45} shows the scalar form factor associated with the $\Gamma_5$ triplet (a), and the $\Gamma_4$ triplet (b).
It is seen that $\Gamma_5$ has a pattern dominated by a hexadecapole, 
while the patterns of 
$\Gamma_3$ and $\Gamma_4$ are dominated by a hexacontatetrapole. 
In the case of PrFe$_4$P$_{12}$, we expect the form factor in Fig.\ref{Gamma_1} is realized in the A sublattice, while the form factor in Fig.\ref{Gamma_3}(c) in the B sublattice.
This is the microscopic image of the scalar order in PrFe$_4$P$_{12}$.
On the other hand, in the case of PrRu$_4$P$_{12}$, we expect the form factor in Fig.\ref{Gamma_1} is realized in the A sublattice, while the form factor in Fig.\ref{Gamma_45}(a) in the B sublattice.  

\begin{figure}[t]
\centering
\includegraphics[width=0.22\textwidth]{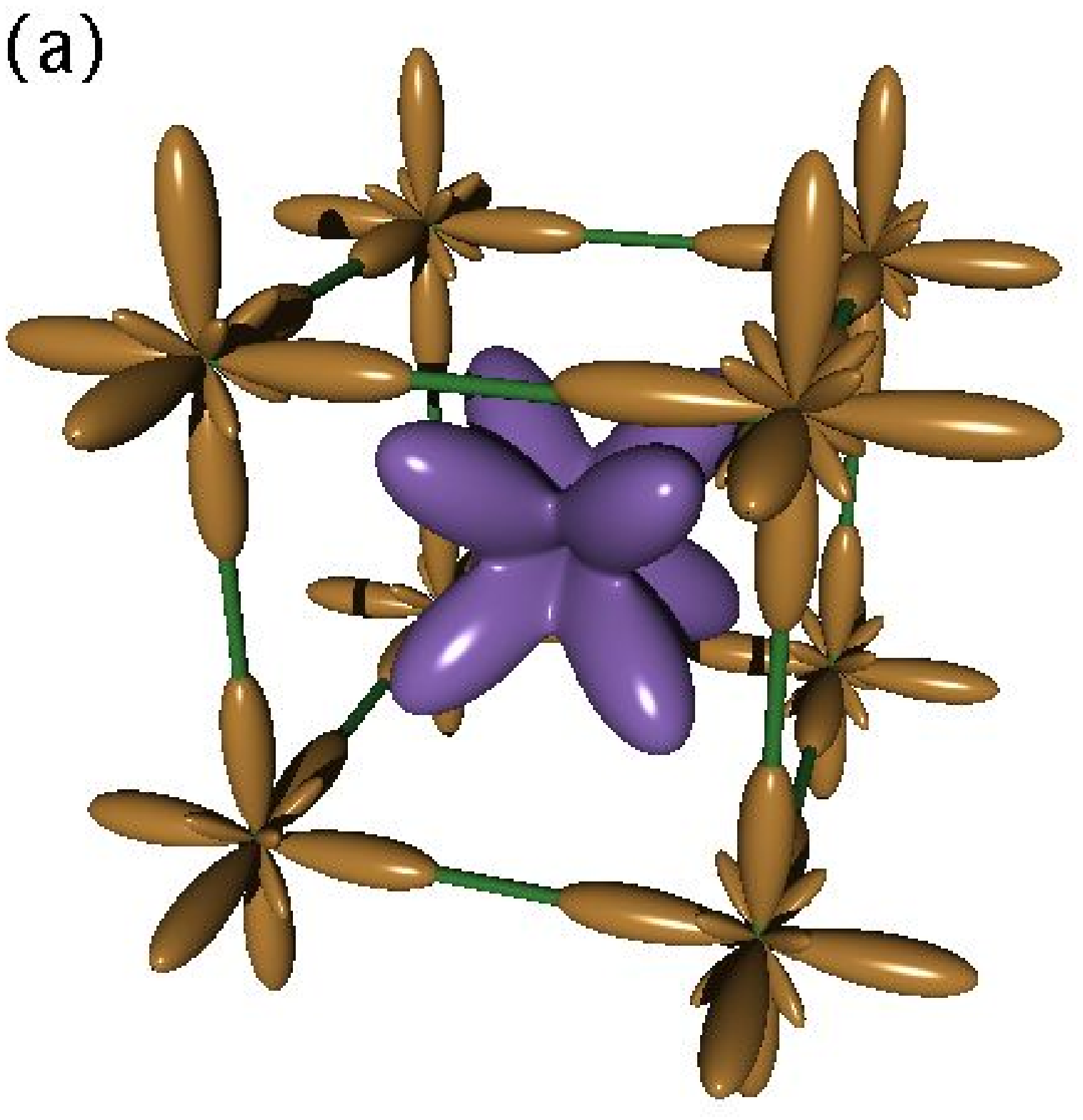}\hspace*{0.2cm}
\includegraphics[width=0.22\textwidth]{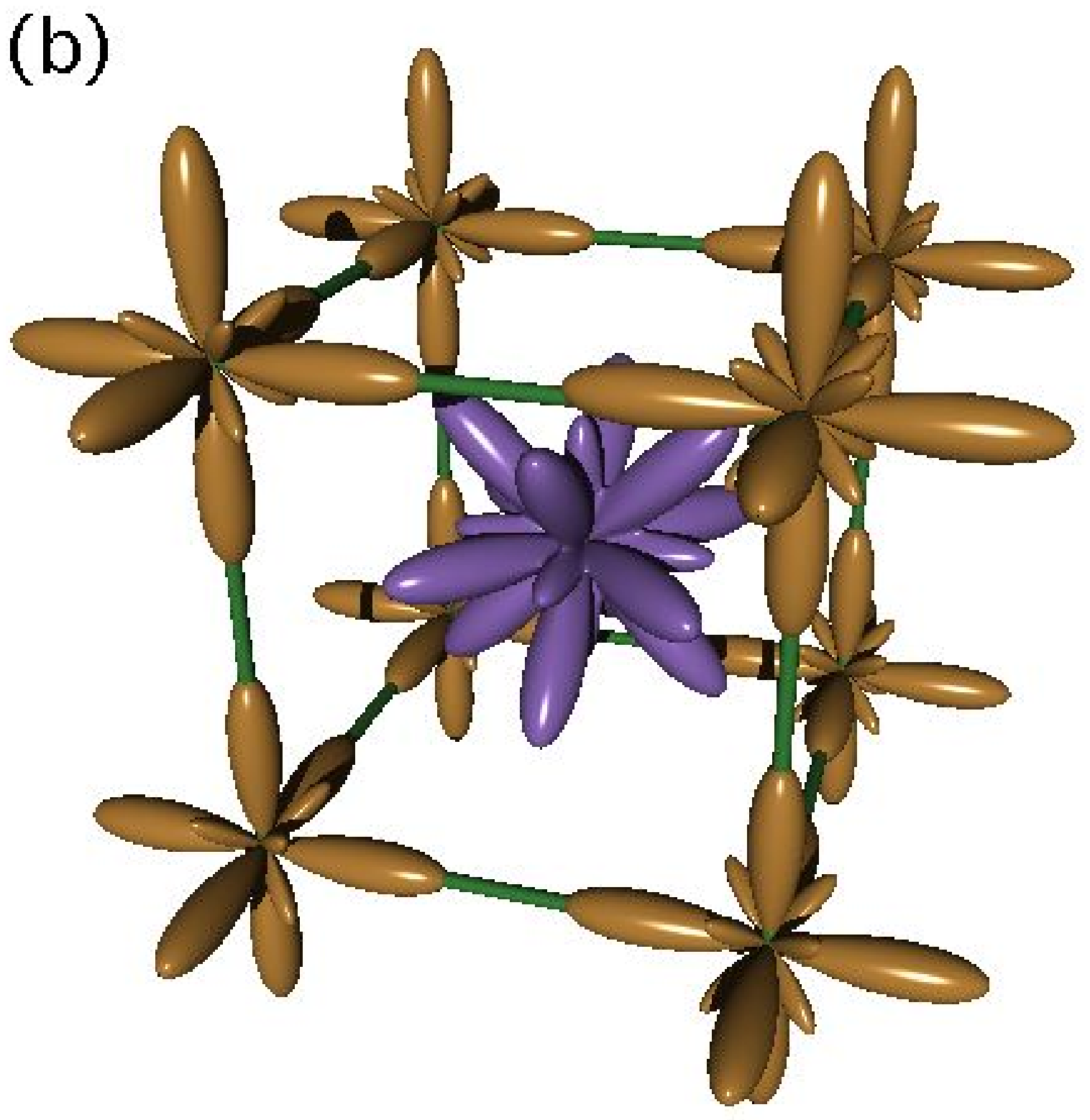}
\caption{
Illustration of the sublattice form factors of the scalar order:
(a) $\Gamma_1$--$\Gamma_{5}$ staggered order; (b) $\Gamma_1$--$\Gamma_{3}$ staggered order.
}\label{lattice}
\end{figure}

Figure \ref{lattice} illustrates the staggered arrangement of angular form factors on the bcc lattice formed by rare-earth ions in skutterudites.  The spatial symmetry in the ordered phase remains cubic, but the unit cell is doubled.  As a result, the superlattice has a simple cubic structure.

As temperature approaches to zero, one of the sublattices has to release the nonzero entropy associated with the doublet or triplet degenerate level. 
In PrRu$_4$P$_{12}$, the triplet seems to split into a singlet and a doublet by about 1 K\cite{aoki2}.  
On the other hand, there is no information about the CEF states in the ordered phase of PrFe$_4$P$_{12}$.
If one of the sublattice is the doublet,  either the lattice distortion or the quadrupolar Kondo effect should break the degeneracy.
Such low-temperature behavior deserves further experimental study.

\section{Landau expansion of the free energy}

We use the phenomenological description of coupling between the scalar order parameter and other degrees of freedom such as magnetization, quadrupole moment, and lattice strain.
In the Landau theory, one expands the free energy in terms of the set of electronic order parameters $\Psi_i$, which are taken to be real.  
Up to fourth-order, we write
\begin{align}
{\cal F}(\Psi) &=
\sum_i\left( 
\frac{1}{2} \alpha_i \Psi_i^2 + 
\frac{1}{4}b_{i}\Psi_i^4  \right)+ 
\sum_{i\neq j} \left( 
g_{ij}\Psi_i^2 \Psi_j +
\frac 12 c_{ij}\Psi_i^2 \Psi_j^2
 \right),
\label{free-energy}
 \end{align}
where 
we have introduced the quantity $\alpha_i = a_i (T-T_i)$.
The constants
$a_i, b_i$ 
are positive, while $g_{ij}$ and $c_{ij}$ can have either sign.
$T_i$ is a hypothetical transition temperature without coupling to other order parameters.   
The actual transition occurs at 
$T_0$ corresponding to the scalar order, which gives the largest of all $T_i$.
For other component $\Psi_i$, we neglect the corresponding $b_i$ in most cases.
These parameterizations have a merit that each coefficient is regarded as a constant as long as the temperature is close to $T_0$, and
the external perturbations are small.  

As explicit constituents of $\Psi_i$, 
we include the scalar order parameter $\psi_{\bf Q}$, the homogeneous magnetization $\mib{M}$, the $\Gamma_3$-type homogeneous quadrupoles.
Furthermore, we also include the lattice strain components $\varepsilon_{xx}$, $\varepsilon_{yy}$, $\varepsilon_{zz}$,
which have a bilinear coupling with quadrupole moments $Q_{ij}$.
The second-order couplings $c_{i\varepsilon}$ with $i=Q,\psi_Q$
can be neglected because the background elastic constant $C_{ij}^{(0)}$ is large enough.
We define the $\Gamma_{3}$ quadrupole moments by
\begin{align}
Q_{u} &= O_{2}^{0}=(1/\sqrt{6})(2J_{z}^2-J_{x}^2-J_y^2), \\
Q_{v}&= O_{2}^{2} = (1/\sqrt{2})(J_{x}^2-J_y^2), 
\end{align}
and introduce the notation $Q^2=Q_{u}^2+Q_{v}^2$.
Similar notations 
$\varepsilon_u, \varepsilon_v $
are used for the strain components. 
Then the magneto-elastic 
couplings are described by
\begin{align}
B\left(  
\varepsilon_u Q_u + \varepsilon_v Q_v
\right)+
g_{Q \varepsilon} Q^2\varepsilon_s 
+ g_{M \varepsilon}\mib{M}^2  \varepsilon_{s} ,
\end{align}
where 
$\varepsilon_s=(1/\sqrt{3})(\varepsilon_{xx}+\varepsilon_{yy}+\varepsilon_{zz})$.
In addition, the free energy in eq.(\ref{free-energy}) includes 
the coupling
\begin{align}
g_{M Q}\left[ \frac{1}{\sqrt{6}} \left(2M_{z}^2 -M_{x}^2-M_{y}^2\right)Q_{u}+\frac{1}{\sqrt{2}} \left(M_{x}^2-M_{y}^2 \right)Q_{v}\right],\nonumber
\end{align}
which we call the tensor coupling.

\section{Magnetic susceptibility}

\subsection{Coupling of moments with the order parameter}

The magnetic susceptibility $\chi$ is obtained from the formula:
$\chi^{-1}= \partial^2{\cal F}/\partial M^2$.
We first consider the case without uniaxial pressure.
For $T>T_0$, we set $\psi_Q=0$, and 
obtain the Curie-Weiss law:
\begin{align}
\chi_{+}^{-1}= a_M (T-T_F),
\label{chi+}
\end{align}
where $T_{F}$ is the Weiss temperature, and the superscript +  indicates $T>T_0$.
In the ordered phase close to the transition temperature, we 
obtain 
\begin{align}
\chi_{-}^{-1} = a_{M}(T-T_{F}) +c_{\psi M} \psi_{\bf Q}^2 
+2g_{M \varepsilon}\varepsilon_{s}.
\label{fullsus}
\end{align}
Since the system is far from the volume collapse,  the last term with $g_{M\varepsilon}$ is almost a constant, and can be incorporated into renormalization of $T_F$.
Using the equilibrium condition $\partial {\cal F}/\partial \psi_{\bf Q} =0$, 
we eliminate $\psi_{\bf Q}^2$ and
obtain the inverse magnetic susceptibility as
\begin{eqnarray}
\chi_{-}^{-1} = 
a_{M}(T-T_{F}) -
c_{\psi M} a_\psi (T-T_0)/
b_{\psi}. \label{chi-}
\end{eqnarray}
There is a peak in the magnetic susceptibility at $T=T_0$ if $\partial \chi_{-}^{-1}/\partial T|_{T_{0}}<0$ is satisfied.
The temperature derivative is calculated as
\begin{eqnarray}
\left. \frac{\partial \chi_{-}^{-1}}{\partial T}\right|_{T_{0}}=
a_{M}-\frac{a_{\psi} c_{\psi M}}{b_\psi} 
\end{eqnarray}
which gives the condition $a_{\psi} c_{\psi M}>a_{M}b_{\psi}$
for occurrence of the peak in $\chi (T)$ at $T=T_0$.
The peak means that the growth of the order parameter gives a negative feedback to the magnetic fluctuation through the repulsive coupling $c_{\psi M}$.

Expressions (\ref{chi+}) and (\ref{chi-})
for the magnetic susceptibility are used to fit the measured result. 
We set $T_{F}=3.5$K for the Weiss temperature \cite{aoki}, 
and value for $a_{M}$ is obtained from the fit in the paramagnetic phase. 
The value for 
the combination $a_{\psi}c_{\psi M}/b_{\psi}$ is obtained from the fit of the susceptibility in the ordered phase. 
The parameters $b_{\psi}$, $a_{\psi}$ and $c_{\psi M}$ are further constrained by the experimental temperature--magnetic field phase boundary.
The fitting with account of these constraints
is shown in the upper panel of Fig.~\ref{fig:4}.

In the case of PrRu$_4$P$_{12}$, the anomaly of $\chi$ at the scalar transition is very small.  
We interpret the smallness in terms of the followings: (i) the system is far from the ferromagnetic instability characterized by $T_F$ in eq. (\ref{chi+}), and
(ii) the coupling constant $c_{\psi M}$ is small.
In fact the elastic anomaly at 
the transition temperature $T_0 \sim 70 $K is less pronounced as compared with PrFe$_4$P$_{12}$ \cite{nakanishi3}.

\subsection{Uniaxial pressure effect}

Intriguing experimental results are obtained for the magnetic susceptibility in the presence of uniaxial stress \cite{saha,matsuda}. Namely, 
the magnetic susceptibility shows large enhancement for the uniaxial pressure applied parallel to the magnetic field direction ($H\|\sigma$), 
while it shows only slight decrease when the pressure is applied perpendicular to the field direction ($H\perp\sigma$).

Now we discuss the properties of the magnetic susceptibility around the transition temperature $T_{0}$.
The direction of the uniaxial stress is taken as
$\sigma\|(001)$, and we consider two different directions of the magnetic field, namely
$H\|(001)$ and $H\|(100)$. 
For small values of the uniaxial stress, it is enough to consider only the linear term in $\sigma$.
Thus,
we obtain the susceptibilities 
$\chi_\parallel$ for $H\|(001)$ and 
$\chi_\perp$ for $H\|(100)$ as 
\begin{align}
\chi_\parallel^{-1} = 
\alpha_M+c_{\psi M}\psi_{\bf Q}^2
+2g_{M\varepsilon}\varepsilon_s
+\frac{4}{\sqrt{6}}g_{MQ}Q_{u},\label{psuspar} \\
\chi_\perp^{-1} = 
\alpha_M+c_{\psi M}\psi_{\bf Q}^2
+2g_{M\varepsilon}\varepsilon_s
-\frac{2}{\sqrt{6}}g_{MQ}Q_{u}.\label{psusperp}
\end{align}
Among the terms appearing in the expression of the magnetic susceptibility, $g_{M\varepsilon}\varepsilon_s$ and $g_{MQ}Q_{u}$ contain the uniaxial stress.
We find in eqs.(\ref{psuspar}) and (\ref{psusperp}) that the former term (scalar) is isotropic, while the latter term (tensor) is anisotropic and has different sign for the two magnetic field directions.
Therefore, the cancellation of the scalar and tensor terms can occur for the case $H\perp\sigma$, which reproduces the experimental situation.
\begin{figure}
\centering
\includegraphics[totalheight=4.5cm,angle=0]{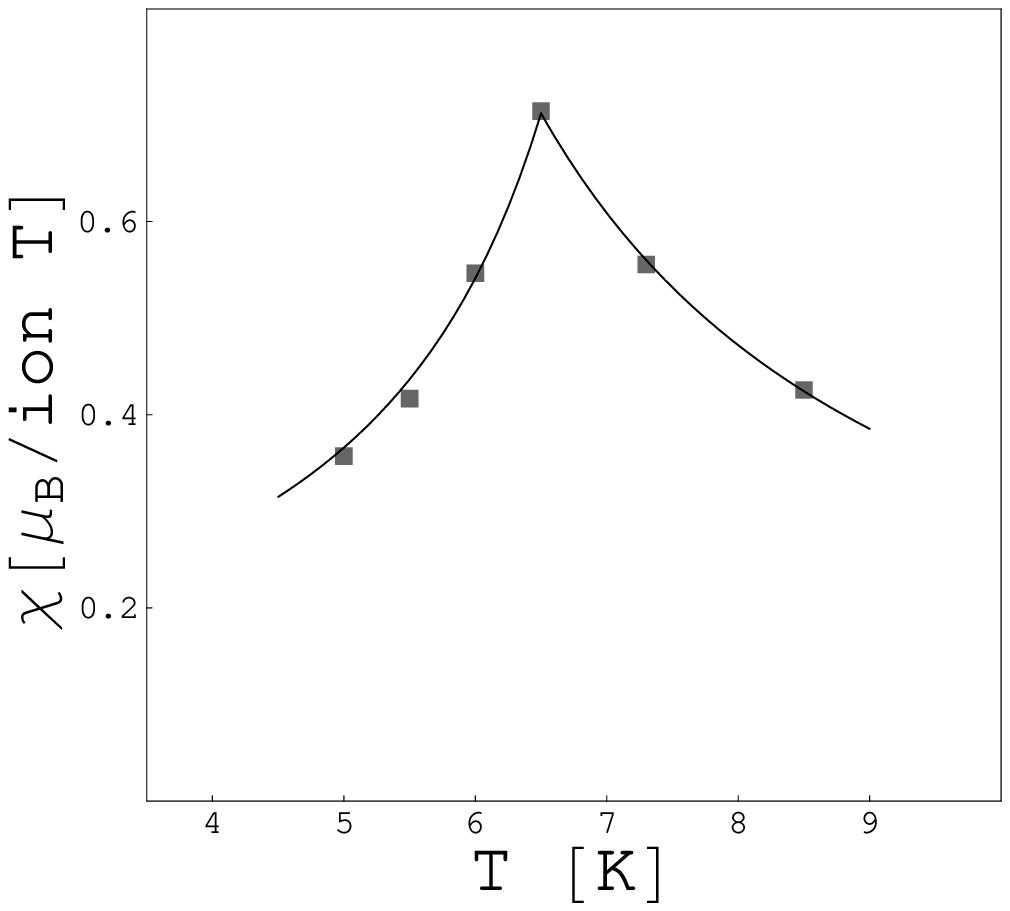}\\
\includegraphics[totalheight=4.5cm,angle=0]{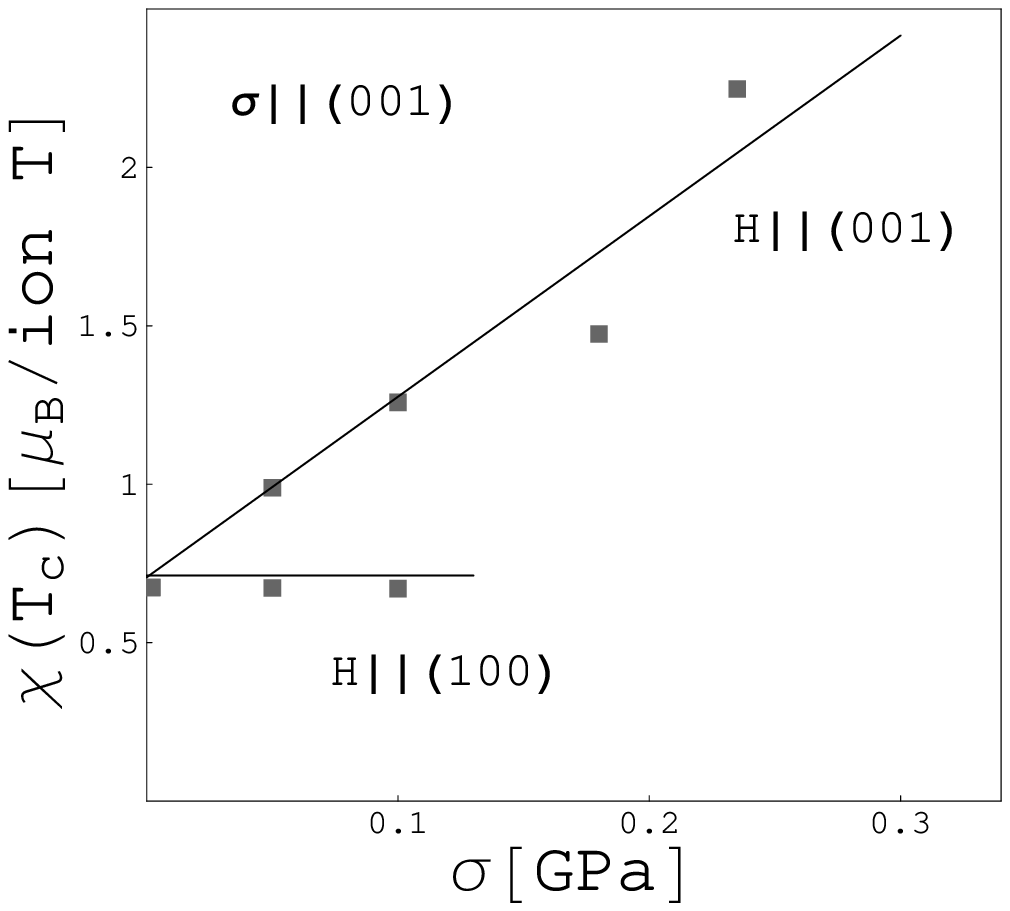}
\caption{{\sl Top:} 
Magnetic susceptibility 
around the transition temperature $T_0$. The parameter values are $b_{\psi}=10^{4}$[Pa], 
$T_{F}=3.5$K,  $a_{M}=9.6\cdot 10^{3}$[Pa$\cdot$$\mu_{B}^{-2}$$\cdot$K$^{-1}$],
$a_{\psi}=1.95\cdot 10^{4}$[Pa$\cdot$K$^{-1}$], $c_{\psi M}=1.37\cdot 10^{4}$[Pa $\cdot$$\mu_{B}^{-2}$].
Boxes represent the measured result taken from ref.~\citen{aoki}.
{\sl Bottom:} 
Boxes represent the measured result taken from ref.~\citen{saha}.
}\label{fig:4}
\end{figure}
Now we set $\psi_{\bf Q}$, $\varepsilon_{s}$ and $Q_{u}$ from the equilibrium conditions 
$\partial {\cal F}/\partial \psi_{\bf Q} =0$, $\partial {\cal F}/\partial \varepsilon_{s} =\sigma_{s}=\sqrt{1/3}\sigma$ and $\partial {\cal F}/\partial \varepsilon_{u} =\sigma_{u}=\sqrt{2/3}\sigma$. 
Keeping only the leading term as in 
calculation without uniaxial stress, 
we obtain the magnetic susceptibilities $\chi_\parallel$ and $\chi_\perp$ in the ordered phase as
\begin{eqnarray}
\chi_{\parallel,-}^{-1} &=&  \chi_{-}^{-1} - \frac{4}{3} \left( \frac{B g_{M Q}}{C_{3}\tilde{\alpha}_{Q}-B^2}\right) \sigma + \frac{2}{\sqrt{3}} \left(\frac{g_{M \varepsilon}}{C_{0}}\right) \sigma,\label{ssuscz}\\
\chi_{\perp,-}^{-1} &=& \chi_{-}^{-1}+ \frac{2}{3} \left(\frac{B g_{M Q}}{C_{3}\tilde{\alpha}_{Q}-B^2} \right) \sigma + \frac{2}{\sqrt{3}} \left(\frac{g_{M \varepsilon}}{C_{0}}\right) \sigma,\label{ssuscx}
\end{eqnarray}
where $\chi_{-}$ is given by eq.(\ref{chi-}). 
Here the relevant elastic constants are written as
$C_3 = C_{11}^{(0)}-C_{12}^{(0)}$, 
$C_0 = C_{11}^{(0)}+2C_{12}^{(0)}$,
and
\begin{eqnarray}
\tilde{\alpha}_{Q}=\alpha_{Q}-
c_{\psi Q}a_\psi (T-T_0)/b_{\psi}.
\end{eqnarray}
The susceptibilities in the paramagnetic phase can be obtained by taking $a_{\psi}=0$ in the eqs.(\ref{ssuscz}) and (\ref{ssuscx}).
Experimentally 
magnetic susceptibility at the transition temperature 
depends linearly on the uniaxial pressure for small values of $\sigma$. 
With a proper choice for the values of $g_{M Q}$ and $g_{M \varepsilon}$, we can fit the measured susceptibilities $\chi_\parallel(T_{c})$ and $\chi_\perp(T_{c})$ at the transition temperature. The result can be seen in the lower panel of Fig.~\ref{fig:4}.
We consider that the almost constant behavior of the magnetic susceptibility $\chi_\perp(T_{c})$ at $T=T_{c}$ is accidental.
Namely, different anisotropic behaviors are also possible depending on the parameters $g_{M \varepsilon}$ and $g_{M Q}$.

\section{Mixing of different angular momenta in CEF states}

The CEF splitting is caused not only by aspherical charge distribution around each rare-earth site, but also by anisotropic hybridization processes \cite{otsuki}.
The hybridization is taken in the form:
\begin{align}
	H_{\text{hyb}} &= 
\sum_{\Gamma \nu \sigma} \left[ V_{\Gamma}
	 p_{\Gamma \nu \sigma}^{\dag} f_{\Gamma \nu \sigma}
	 +\text{H.c.} \right],
\end{align}
where $\Gamma$ is an irreducible representation in $T_h$, and $\nu$ is an element therein.  If there are different electronic states with the same representation, we distinguish them in terms of the index $\alpha$ such as $\Gamma (\alpha)$.

In the standard theory for rare-earth,  the CEF states are obtained by 
diagonalizing the Hund's rule ground states with given $J$ under the CEF potential.  
The mixing of excited states with different $J$ is neglected because the spin-orbit splitting is much larger than the typical CEF splitting.  
However, the difference of the energy splittings alone does not justify the neglect of higher $J$ because the degree of mixing is not only determined by the Coulombic CEF potential, but also by the covalent hybridization with much larger energy scale.
In order to clarify the situation, we use the Brillouin-Wigner perturbation theory which gives the formally exact series of perturbed wave function $\Psi$ in terms of unperturbed on $\Phi$ as follows:
\begin{align}
\Psi = \Phi + \frac{\cal Q}{E-H_0}H_{\text{hyb}} \Phi+\left( 
\frac{\cal Q}{E-H_0}H_{\text{hyb}}  \right)^2\Phi+\ldots,
\end{align}
where $\cal Q$ is the projection operator to make states orthogonal to $\Phi$, and $E$ is the exact energy of $\Psi$.
The unperturbed Hamiltonian $H_0$ describes the decoupled f-electron and ligand states.  Then the intermediate state $H_{\text{hyb}} \Phi$ is dominated by $4f^{n-1}$ ($4f^{n+1}$ )
states plus an extra ligand electron (hole).  
In the following 
we take the specific case where the trivalent Sm 
with $4f^5$ configuration is the ground state, which
has dominant intermediate states with $4f^6$.
The weight of the $O(H_{\rm hyb})$ term in $\Psi$ is small because the hybridization is less than 1 eV, and the excitation energy to Sm$^{2+}$ is typically a few eV.

\begin{figure}
\centering
\includegraphics[width=0.38\textwidth]{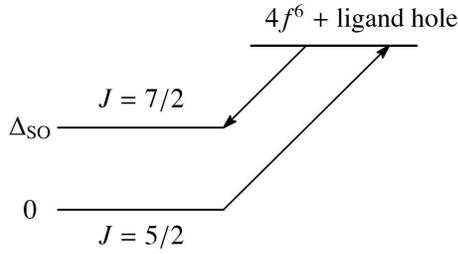}
\caption{
Second-order perturbation processes to mix $J=7/2$ states in the CEF ground state.  The spin-orbit splitting $\Delta_{\rm SO}$ is about 0.12 eV in Sm$^{3+}$.
}
\label{2nd-mixing}
\end{figure}

On the other hand, some the $O(H_{\rm hyb}^2)$ term has a larger weight in $\Psi$ than the first-order term.  
This is because the second intermediate states are
dominated by the Sm$^{3+}$ configuration which has the same $4f$ electron number as the ground state but with different values of $J$.  
Then the excitation energy in $E-H_0$ is of the order of spin-orbit splitting, which is especially small in the case of $4f^5$ with merely 0.12 eV.  
If we put $H_{\rm hyb} \sim 0.3$eV and $U_{\rm eff}\sim E-H_0\sim 2$eV, we can roughly estimate as $H_{\rm hyb}^2/(U_{\rm eff} \Delta_{\rm SO})\sim 0.4$.
Hence the ratio of hybridization over the excitation energy exceeds unity, and the weight of mixed states with $J=7/2$ or higher is not negligible.  
Figure \ref{2nd-mixing} illustrates the perturbation processes.

The $\Gamma_{67}$ ground CEF states have wave functions with $J=5/2$:
\begin{align}
|a\pm\rangle &= \sqrt{5/6}|\pm 5/2\rangle +\sqrt{1/6}|\mp 3/2\rangle, \\
|b\pm\rangle &= |\pm 1/2\rangle,
\end{align}
where $a$ and $b$ specify the orbital quantum number, and $\pm$ the Kramers partners.
In the case of $J=7/2$, the wave functions with $\Gamma_{67}$ symmetry are given by
\begin{align}
|\alpha\pm\rangle &= \sqrt{1/4}|\pm 5/2\rangle 
+\sqrt{3/4}|\mp 3/2\rangle, \\
|\beta\pm\rangle &= 
\sqrt{7/12}|\pm 7/2\rangle -\sqrt{5/12}|\mp 1/2\rangle,
\end{align}
where the orbitals are now specified by $\alpha, \beta$.
Any linear combination of a $J=5/2$ state in $\Gamma_{67}$ and that with $J=7/2$ satisfies the symmetry requirement for the $T_h$ group.  Hence the eigen function of the actual CEF potential is determined by the mixing term connecting $J=5/2$ and 7/2 manifolds.
According to our estimate in the preceding paragraph, the magnitude of the effective mixing potential
\begin{align}
V_{\rm eff}= H_{\text{hyb}}\left( \frac{\cal Q}{E-H_0}H_{\text{hyb}}  \right)^2
\end{align}
should be larger than, or at least of the same order of magnitude as
the spin-orbit splitting $\Delta_{\rm SO}\sim 0.12$ eV.
If the Sm$^{3+}$ state is closer to mixed valence, the excitation energy to $4f^6$ (Sm$^{2+}$) should also be small.
As a result, the CEF ground state is expected to have considerable weight of $J=7/2$ and higher angular momenta.
It should be interesting to estimate the mixing more quantitatively.

\section{Bilinear coupling of dipoles and octupoles in SmRu$_4$P$_{12}$}

The most important consequence of the mixing of $J=7/2$ and even larger angular momentum in the $\Gamma_{67}$ CEF state is that the matrix element of the sixth-rank tensor $O_6^t=O_6^2-O_6^6$ becomes nonzero.
Then the symmetry $T_h$ lower than $O_h$ makes mixing of two triplet representations $\Gamma_4$ and $\Gamma_5$, which has been recognized important in Pr skutterudites.
In the $T_h$ symmetry, there is a single triplet representation called $\Gamma_4$, and different states are specified by a superscript such as 
$\Gamma_4^{(1)}$ or $\Gamma_4^{(2)}$.
Physically speaking, most dipoles and octupoles mix in the $T_h$ group.  The only pure octupole is the pseudo-scalar $\Gamma_{1u}$ which transforms as a third-rank tensor $xyz$.  
In $O_h$, 
the odd representation $\Gamma_{5u}$, which transforms as $x(y^2-z^2)$  and its cyclic partners, is also a pure octupole.

It has been suggested by Yoshizawa {\it et al.}\cite{yoshizawa1} that the ordered phase in SmRu$_4$P$_{12}$ has the dominant $\Gamma_{5u}$ octupole component, which mixes with the dipole component $\Gamma_{4u}$  in the notation of $O_h$.  Yoshizawa's idea is most simply illustrated by the following form of the Landau free energy:
\begin{align}
{\cal F} = 
\frac 12 a_4 (T-T_4)\Psi_4^2 +
\frac 12 a_5 (T-T_5)\Psi_5^2 +v\Psi_4\Psi_5,
\label{bilinear}
\end{align}
with neglect of higher order terms.
The last term with a real coupling constant $v$ represents the mixing between dipoles and octupoles
represented by $\Psi_4$ and $\Psi_5$, respectively.
Without the coupling term, each order would have set in at temperature $T_4$ or $T_5$.
The actual transition temperature is given by
\begin{align}
T_{c+} = \frac 12\left( T_4+T_5 \right)
+\left[ \frac 14\left( T_4-T_5 \right)^2 +\frac{v^2}{a_4a_5}\right]^{1/2} ,
 \label{T_c+}
\end{align}
while the partner temperature $T_{c-}$ with the negative sign for the square root in eq.(\ref{T_c+}) is not a true transition, since the order parameter already grows below $T_{c+}$.  If the coupling $v$ is small, however, $T_{c-}$ may appear as a crossover with some structure in physical quantities.

This scenario explains most naturally the appearance of the internal magnetic field \cite{hachitani,tsutsui,higemoto}, ferro-quadrupole moment \cite{mito}, and the large elastic anomaly \cite{yoshizawa2,sun}.
These features are analogous to those in the phase IV of Ce$_{1-x}$La$_x$B$_6$ with $x\sim 0.7$, where the antiferro octupole order has been proposed \cite{kusunose,kubo}, and confirmed by various measurements \cite{goto,sakakibara2,mannix}.
The cubic symmetry in Ce$_{1-x}$La$_x$B$_6$ allows pure octupoles which should have zero internal field at the Ce nucleus site, but with finite off-center field which has recently be detected by neutron scattering \cite{kuwahara}.

\begin{figure}
\centering
\includegraphics[width=0.45\textwidth]{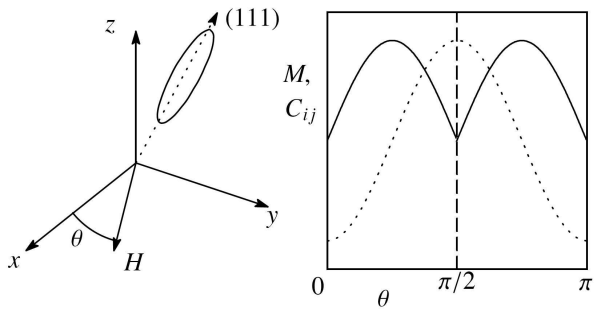}
\vspace{0.3cm}
\caption{
Preferred quadrupole domain for magnetic field in the 
$xy$-plane with $0< \theta <\pi/2$ {\it (Left)}.
In magnetization and elastic constants in SmRu$_4$P$_{12}$, a domain switching to $[1\bar{1}1]$ for $\pi/2<\theta<\pi$ 
should lead to a cusp structure on top of the two-fold pattern shown by the dotted line {\it (Right).}
}
\label{domain}
\end{figure}

The domain structure of SmRu$_4$P$_{12}$ which is consistent with the NQR result\cite{mito} is the same as that in Ce$_{1-x}$La$_x$B$_6$;  there appear ferroquadupole domains with principal axis along [111]  and other three equivalent directions.   
If a magnetic field favors a quadrupole whose principal longer axis is the closest to the field direction, 
a sudden switching from one domain to another should take place as magnetic field is rotated.  This switching should appear as a cusp structure in physical quantities such as magnetization and elastic constant.  In fact, Yoshizawa {\it et al.} have recently observed an intriguing pattern in the elastic constant of SmRu$_4$P$_{12}$ \cite{yoshizawa07}.  Figure \ref{domain} 
illustrates our interpretation of the pattern.

Experimentally, the lower transition becomes more and more visible as the applied magnetic field becomes stronger.  In the above scenario, it seems difficult to reproduce such a behavior as long as the coupling $v$ is independent of magnetic field.  
Another possibility is that the second transition is also a real phase transition above 
a critical magnetic field, but becomes a crossover below the critical field.
Then the presence of critical point at finite temperature and field requires another model which is certainly more complicated than the one given by eq.(\ref{bilinear}).  Within the Landau phenomenology, we have been unable to find a model with the desired property; the second transition, if real, always starts from zero temperature instead of a critical point at finite temperature.
It remains a challenge for theory to construct a microscopic model which goes beyond the scope of the Landau phenomenology.
From experimental side, it is desirable 
to decide whether the second transition is a sharp crossover, or a real transition.  

\section*{Acknowledgments}

We are grateful to M. Yoshizawa and Y. Aoki
for sharing their experimental results prior to publication,
and to Prof. H.  Harima for discussion about the CEF eigenstates.

\end{document}